\documentclass[final]{elsarticle}

\usepackage{lineno,hyperref}
\modulolinenumbers[6]

\usepackage{natbib}
\usepackage{graphicx}
\usepackage{float}
\usepackage{bbold}
\usepackage{amsmath}
\usepackage{amsfonts}
\usepackage{amssymb}
\usepackage{mathrsfs}
\usepackage{mattens}
\usepackage{amsthm}
\usepackage{upgreek}
\usepackage{textcomp}
\usepackage{placeins}
\usepackage{subcaption}
\usepackage{tikz}

\usepackage{soul}
\usepackage{todonotes}

\usepackage{xcolor}
\usepackage{framed}
\colorlet{shadecolor}{blue!20}

\usepackage{listings}

\let\vec\mathbf
\let\mat\boldsymbol

\journal{Journal of the Mechanics and Physics of Solids}

\begin{document}

\begin{frontmatter}

\title{Contact phase-field modeling for chemo-mechanical degradation processes. Part II: Numerical applications with focus on pressure solution.}


\author[mymainaddress]{A. Gu\'evel}
\corref{mycorrespondingauthor}
\cortext[mycorrespondingauthor]{Corresponding author}
\ead{alexandre.guevel@duke.edu}

\author[mymainaddress]{H. Rattez}


\author[mymainaddress]{E. Veveakis}

\address[mymainaddress]{Duke University, School of Civil and Environmental Engineering, USA}

\begin{abstract}

The microstructural geometry (MG) of materials has a significant influence on their macroscopic response, all the more when the process is essentially microscopic as for microstructural degradation processes. However, the MG tends to be approximated by ideal spherical packings with constitutive description of the microstructural contacts. Interfaces tracking models like phase-field modeling (PFM) are promising candidates to capture the microstructures dynamics. Contact PFM (CPFM) enables to include catalyzing/inhibiting (CI) effects, accelerating/delaying equilibrium, such as temperature or the presence of certain constituents. To emphasize the influence of geometry and CI effects, we study numerically the chemo-mechanical response of digitalized geomaterials at the grain scale. An application to pressure solution creep (PSC) shows the importance of the MG and how the influence of temperature and clay can be taken into account without explicit modeling. As already inferred in previous works on PSC, the lack of MG considerations could be the reason why a unique description of PSC is missing. A simple reason could be that PSC is directly dependent on the strain concentration, which is directly dependent on the MG. This is our motivation here to investigate and suggest the influence of the MG on a degradation process like PSC.

\end{abstract}


\end{frontmatter}



\section{Introduction} 

\subsection{Summary of our model's theoretical foundations}

We have set forth in the first part of this work \cite{Guevel2019a} the theoretical foundations of CPFM. This is an extended PFM based on a non-equilibrium thermodynamic framework, contact thermodynamics, incorporating for now chemo-mechanical coupling. The mechanical effect is based on elasticity triggering the production of weak phase, similarly to dissolution. The chemical effect allows the opposite reaction, the production of strong phase in the zones away from the ones with large mechanical loading. The precise discrimination of the mechanical response within the system is ensured by the PFM capturing the actual interfaces. For this reason, we will apply this model in the present second part to microstructures with complex geometries, those of geomaterials. 

The novelty of our extended PFM resides in the term $\mu \Delta\dot\phi$ ($\phi$ is the order parameter), added to the usual term $\dot\phi$. We claim that the latter characterizes the normal variations of the interfaces curvature and the former their change of orientations i.e. tangential variations. Thus $\mu$, that we call PFM viscosity, quantifies the resistance for a rough geometry to smoothen. In that sense, $\mu$ encapsulates the kinetics of microstructural changes and could be described with the different activation energies of the CI effects associated to the main process, such as temperature. 

The influence of the main ingredients of our model, the bulk energy input $\chi$ and the PFM viscosity $\mu$ are first benchmarked. Then, in order to thoroughly exhibit our model's capabilities, we numerically model PSC at the grain scale. The geomaterials' MG is input in our model via the digitalization of micro CT-scans. Let us introduce now the main facets of PSC, in particular where the influence of MG stands in the literature.

\subsection{Pressure solution creep}


The concept of PSC finds its origin in the realization that the deformation of materials in the presence of fluids cannot accounted for solely on mechanical bases, but combined with chemical effects. Fluid-bearing materials like geomaterials fall into this consideration. The term "pressure solution" was originally coined by the geologist Sorby, in application to geomaterials, for which processes "mechanical force is resolved into chemical action" \cite{Sorby1863}.


PSC is a major factor in the dynamics of Earth's crust, mainly in the upper crust, as in lithogenesis (\cite{Heald1956}), tectonic (\cite{Schwarz1996}, fault reactivation and earthquakes (\cite{Sleep1992}, \cite{Renard2000}). An outstanding observable example of PSC is the formation of stylolites (see for instance the review in \cite{Gratier2013a}). Finally, PSC can play a role in underground storage of nuclear waste both in the case of rock salt cavities \cite{Urai1986} and bentonite buffer that could make the metallic container sink \cite{Shin2017}. 



The process happening to geomaterials' grains can be compared to the one happening to salt or sugar in humid air, where grains tend to stick together under sufficient amount of loading (due to gravity) and humidity.
PSC is usually the first degradation process to occur, provided the system is given enough time, at least before plasticity and breakage \cite{DeBoer1977}. Experimental results corroborate indeed that diagenesis takes place chiefly by pressure solution rather than by grain straining or crushing (see \cite{Lowry1956, Renton1969, Schwarz1996} e.g.). It is thus primordial to take PSC into account when there are inter-granular fluids, inasmuch as it can predetermine other processes.

Pressure solution creep (PSC) is a serial stress-driven mass transfer process that can be described in four stages: (1)dissolution at grain impingements, (2)diffusion of the solute out of the contact, (3)deposition/precipitation on the pore surface and potential (4)diffusion of the solute to other pores (see fig.1 below adapted from \cite{Gundersen2002}).

\begin{figure}[h!]
\centering
\includegraphics[scale=0.4]{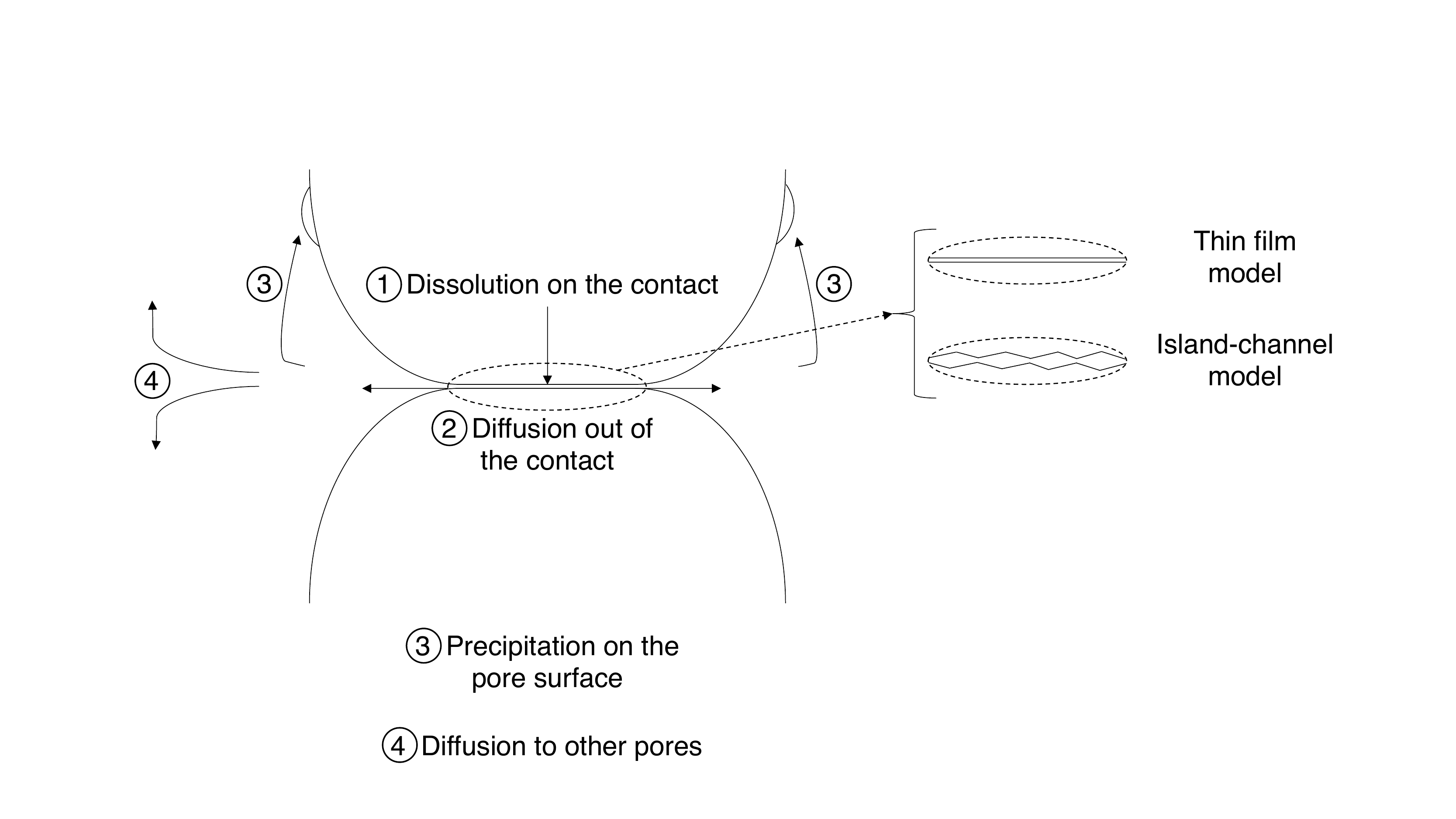}
\caption{4 main PSC processes and 2 main models for grain contacts}
\end{figure}

 The corresponding indicators to look for in experiments are mainly interpenetration and overgrowth. The main drive is mechanical (constant) loading, triggering dissolution in stressed regions and allowing precipitation in sheltered regions relatively unstressed (after solute diffusion), first experimentally inferred by Griggs \cite{Griggs1940}. Since it is an en-serie process, the rate of the slower process will govern the rate of the overall process. Different limiting processes correspond to different creep laws. 

A key process is the diffusion through the compressed thin layer of trapped fluid between the grains triggering the dissolution. Two main models have been used to explain it: the fluid layer model \cite{Weyl1959, Renard1997} and the island-channel network model \cite{Raj1981}. Note that the diffusivity of the trapped fluid is much lower than the diffusivity of the free fluid in the pores or fractures. The main factors influencing the PSC are the stress, the temperature, the geometry (grain size e.g.), the porosity and the diffusion (and advection if any) coefficients.\\


We believe that the existing PSC models present two main limitations. Firstly, there is no clear consensus as for the rate-limiting process (i.e. the slowest process that will control the overall creep law) \cite{Croize2013}. Therefore PSC cannot be described presently with a unique law \cite{Raj1982, Gratier2013a}. Secondly, the actual MG is typically averaged by regular packs of mono-dispersed spherical particles. This approximation neglects the highly irregular geomaterials' MG and its continuous variations during compaction. Even though the grain boundary structures can be modeled and taken into account in the creep laws by assuming a thin-film model or islands-channels model, it seems that the grain size distribution can be misleading. This can cause major discrepancies between model and experimental \cite{Niemeijer2009,Croize2013}. The variation of grain distribution of the actual MG can potentially localize the deformations and lead to spatio-temporal variations in pressure solution rates. Thus overlooking the actual MG can mean passing over potential microscopic instabilities (at the grain scale) that can create major disruptive macroscopic events like earthquake ruptures \cite{Niemeijer2009}. It is clear that those two limitations are closely related insofar as PSC is essentially a grain-scale process and therefore characterizing it amounts to adequately modeling the MG. 
In a first approach, we will address this impediment by modeling the MG with our CPFM.


\section{Problem set-up}

Let us remind the problem's equation for PFM with chemo-mechanical coupling in dimensionless form:

\begin{equation}
\begin{cases} 
-\mu\Delta\dot{\phi} + \dot{\phi} = \alpha\Delta\phi - f(\phi,\mat\epsilon,c) \\ 
\tau^*\dot{c} = D^*\Delta{c} - \tau^*\dot\phi - \hat\beta^*(\phi)
\end{cases}
\end{equation}

With $\mu=\frac{\tau_2}{\tau_1 l_0^2}$ (phase-field viscosity), $\alpha=\frac{\Gamma}{G l_0^2}$, $f(\phi,\mat\epsilon,c)=g'(\phi)+(\chi(\mat\epsilon)-\beta^* c)h'(\phi)=g'(\phi)+\hat\chi(\mat\epsilon,c)h'(\phi)$, $\chi(\mat\epsilon) = \frac{1}{2}\mat\epsilon.(\mat{C_B}-\mat{C_A})\mat\epsilon$, $\tau^*=\tau_3/\tau_1$, $D^*=\frac{D}{G l_0^2}$, $\hat\beta^*(\phi) = \frac{\beta}{G} \left[\frac{\int_V(1-h(\phi))dV}{\int_V dV}+1-h(\phi) \right]$

It is reminded that the mechanics is imply solved as follows:

\begin{equation}
\begin{cases} 
\nabla.\mat\sigma=0 \\ 
\mat\sigma = \overline{\mat{C}}(\phi)\mat{\epsilon} + \tau_3 \mat{\dot{\mat\epsilon}}
\end{cases}
\end{equation}

With $\overline{\mat{C}}(\phi)= \mat{C}^A (1-h(\phi)) + \mat{C}^B h(\phi)$ the homogenized elastic tensor. $A$ is the weak phase ($\phi=0$( and $v$ the strong phase ($\phi=1$.
For the sake of simplicity, we assume in the present work that $\tau_3=0$. We will work in 2D plane strain unless mentioned otherwise, then the elastic energy of phase $K$ reads:

\begin{equation}
H_K=\frac{1}{2}\lambda_K({\epsilon_{xx}^{K}}^2+{\epsilon_{yy}^{K}}^2)+\mu_K({\epsilon_{xx}^{K}}^2+{\epsilon_{yy}^{K}}^2+2{\epsilon_{xy}^{K}}^2)
\end{equation}

With $\lambda_K$ and $\mu_K$ the Lam\'e parameters of phase $K$.

The reference time scale is fixed to $t_0=\frac{\tau_1}{G}$ corresponding to the relaxation of the normal variations of the interface. The reference length scale $l_0$ shall be fixed with respect to the problem's dimensions at stake.
All the model's details can be found in \cite{Guevel2019a}. From the linear stability analysis therein, one should choose $\alpha$ significantly smaller than the problem's dimension, so that only the two phases at stake are stable (and observed). That will also discriminate the diffusion of the solute with respect to the interface diffusion ($D^* \gg \alpha$). Different reaction diffusivities are characteristic of reaction-diffusion systems. We will see that in practice it not always practical to achieve.\\

%

\section{Numerical implementation and benchmarks}

In the following numerical examples, we intend to illustrate the roles played by the characteristic parameters $\chi$ and $\mu$, and more generally how including the MG impacts the material's response. It seems that $\chi$ plays the role of phase change catalyst whereas $\mu$ plays the role of a phase change inhibitor. Note however that $\chi$ encapsulates a static effect but $\mu$ a dynamic effect. The former is illustrated by displaying weak phase nucleation and the latter by considering geometrical effects.

\subsection{Multiphysics Object-Oriented Simulation Environment}

The Multiphysics Object Oriented Simulation Environment
(MOOSE \footnote{http://mooseframework.org}) used here is a finite-element-based code dedicated to multiphysics nonlinear problems. Its efficiency comes, inter alia, from the facts that it is object-oriented, hence easily reusable, fully coupled, fully implicit, automatically parallel, it uses unstructured meshes, and more practically it is open source and thus makes the most of a large and helpful users community.\\

Following \cite{Tonks2012}, the system is discretized in time using the Jacobian-Free Newton Krylov method (JFNK) and in space with a FEM formulation. The equation is solved by minimizing its residual written in the weak form the lastly obtained dimensionless CPFM. The weighted integral residual projection reads:

\begin{equation}
-\left(\mu\Delta\dot{\phi},\varphi_m \right) + \left(\dot{\phi},\varphi_m \right) - \left(\Delta\phi,\varphi_m \right) + \left(f_\chi(\phi,\mat\epsilon),\varphi_m \right)=0
\end{equation}

Where $\varphi_m$ is a test function, and $\left(.,. \right)$ is the usual integral inner product. Integrating by parts the first and third term, noting $\left<.,.\right>$ the boundary terms yield:

\begin{equation}
\left(\mu\nabla\dot{\phi},\nabla\varphi_m \right) - \left<\mu\nabla\dot{\phi}.\vec{n},\varphi_m \right>+ \left(\dot{\phi},\varphi_m \right) + \left(\nabla\phi,\nabla\varphi_m \right) - \left<\nabla\phi.\vec{n},\varphi_m \right>  + \left(f_\chi(\phi,\mat\epsilon),\varphi_m \right)=0
\end{equation}

Likewise, the macro-force balance $\nabla.\mat\sigma=0$ is implemented using the weak form:

\begin{equation}
\left(\mat\sigma,\nabla\varphi_m \right) - \left<\mat\sigma.\vec{n},\varphi_m \right> = 0
\end{equation}

Each term $\left(.,. \right)$ inherits from a usually pre-implemented C++ base class called kernel in MOOSE. The existing kernels have been customized to fit our model.

\subsection{Role of $\chi$: phase change activation}

We first focus on the influence of $\chi$, characterizing the source energy for phase change, in our case elastic energy. The system's energy input is to destabilize its stable double-well organization and therefore to initiate phase changes. This can be visualized with the graph of the bulk energy term $B(\phi,\epsilon)=Gg(\phi)+\bar{H}(\epsilon,\phi)$ with $\bar{H}(\epsilon,\phi)=H_B(\epsilon)h(\phi) + H_A(\epsilon)(1-h(\phi))$. We consider the phase A to be the weak phase and B the strong phase, meaning A is much more deformable than B. Note that is consistent with Landau's theory to take an order parameter increasing with the (microscopic) "order" of the material considered. Here and in the next examples $\phi=0$ corresponds physically to pores (filled with liquid or not) and $\phi=1$ to a solid state.  If the energy input $\bar{H}(\epsilon,\phi)$ is low enough, the system has a stable double-well organization (blue graph). If $\bar{H}(\epsilon,\phi)$ is high enough, the double well is tilted, which favors the least energy-demanding organization, i.e. an expansion of phase A at the expense of phase B (orange graph). This destabilization has been determined in \cite{Guevel2019a} to be any value of $\chi(\mat\epsilon) = \frac{1}{2}\mat\epsilon.(\mat{C_B}-\mat{C_A})\mat\epsilon$ larger than $\chi_0 \approx 1/3$ (with the proviso that the perturbation characteristic length is much higher than the interface characteristic width).

\begin{figure}[h!]
\centering
\includegraphics[scale=0.7]{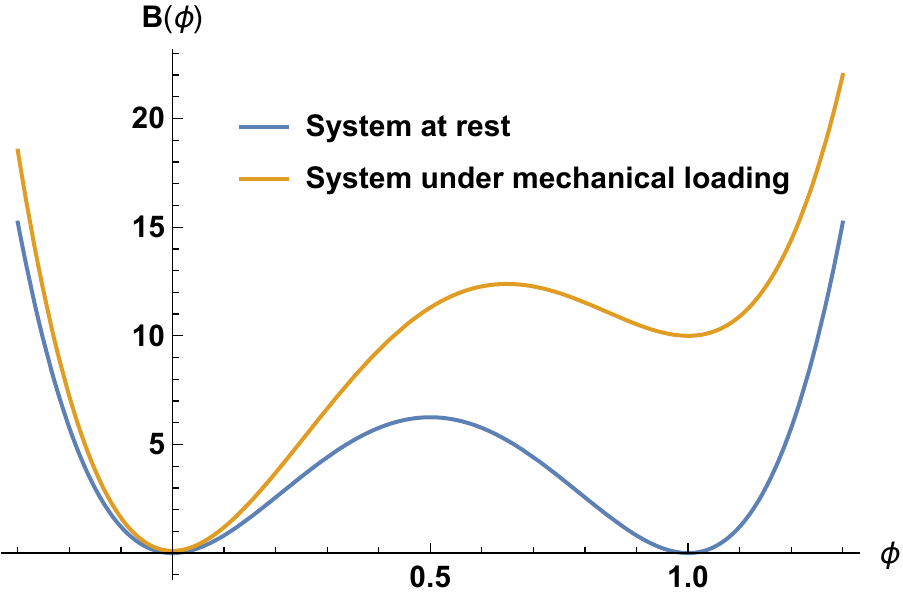}
\caption{Graph of the double-well potential $Gg(\phi) \approx B(\phi)$ with $G=10$ in blue and of the tilted double-well potential $B(\phi,\epsilon)=Gg(\phi)+\bar{H}(\epsilon,\phi)$ with $G=10$, $H_A(\epsilon)=0.1$ and $H_B(\epsilon)=10$ in orange}
\end{figure}

Thus we want to observe the nucleation of the phase A under mechanical loading. Let us consider the extreme case where there is no phase A at beginning, with initial conditions randomly fluctuating in the phase B, i.e. $\phi \in [0.79,1]$. We remind that, as explained in\cite{Guevel2019a}, we consider the pure phases A and B for respectively $\phi \in [0,0.21]$ and $\phi \in [0.79,1]$, while the interface corresponds to the spinodal interval $\phi \in [0.21,0.79]$. 
The boundary conditions are those of a uniform compression, i.e. constantly moving boundaries towards the center, and null Neumann conditions for the order parameter. We choose (displacement-controlled) loading directly proportional to the simulation time ($1*t$). Now let us set up some numerical values for the parameters. In all our numerical studies, we will work with the following consistent units system: length in $mm$, mass in $kg$, time in $ks$, energy in $J$, pressure/stress in $GPa$. The dimensionless interfacial coefficient $\alpha=\frac{\gamma l_i}{G l_0^2}$ can be estimated by choosing a surface tension $\gamma=0.1 Pa.m = 10^{-7}J/mm^2$ (for a solid-fluid interface, cf Leroy2001 e.g.), a interface width $l_i$ of $1nm=10^{-3}mm$, a problem's length scale $l_0$ of $1mm$ and a double-well barrier $G$ of $1 J/mm^3$; then $\alpha=10^{-12}$. As often in PFM, $\alpha$ is an "epsilon" term. Yet, for numerical purposes, this value is chosen as small as possible to maintain good convergence. We choose $\alpha=0.01$ in the present case. The mechanical properties are chosen to consider a solid (strong phase B) containing porous fluid (weak phase A). As a first approximation, we model the pores phase as a solid as well, as in \cite{Kassner2001}. As such, the pores fluid (air and/or liquid) will be then taken as a shear-free solid much more deformable than the matrix phase. Thus we choose for instance the  $\lambda_A=1 GPa$, $\mu_A=0$, $\lambda_B=\mu_B=30 GPa$ ($\lambda_K$ and $\mu_K$ being the Lam\'e's first and second parameters of the phase $K$ respectively). We keep $\mu=0$ for now (not to be confused with the second Lam\'e parameter). Finally, we choose the reference length $l_0=1mm$ (side of the initial square). We observe below the initial, softening and final stages of our simulation for two different different values of $\chi$ (a function of the strain and the Lam\'e parameters). Associated is the stress measured at the top surface (averaged) vs the vertical displacement or shortening (vs). All the following stress/displacement curves will be given similarly.


\begin{figure}[h!]
\centering
\includegraphics[scale=0.4]{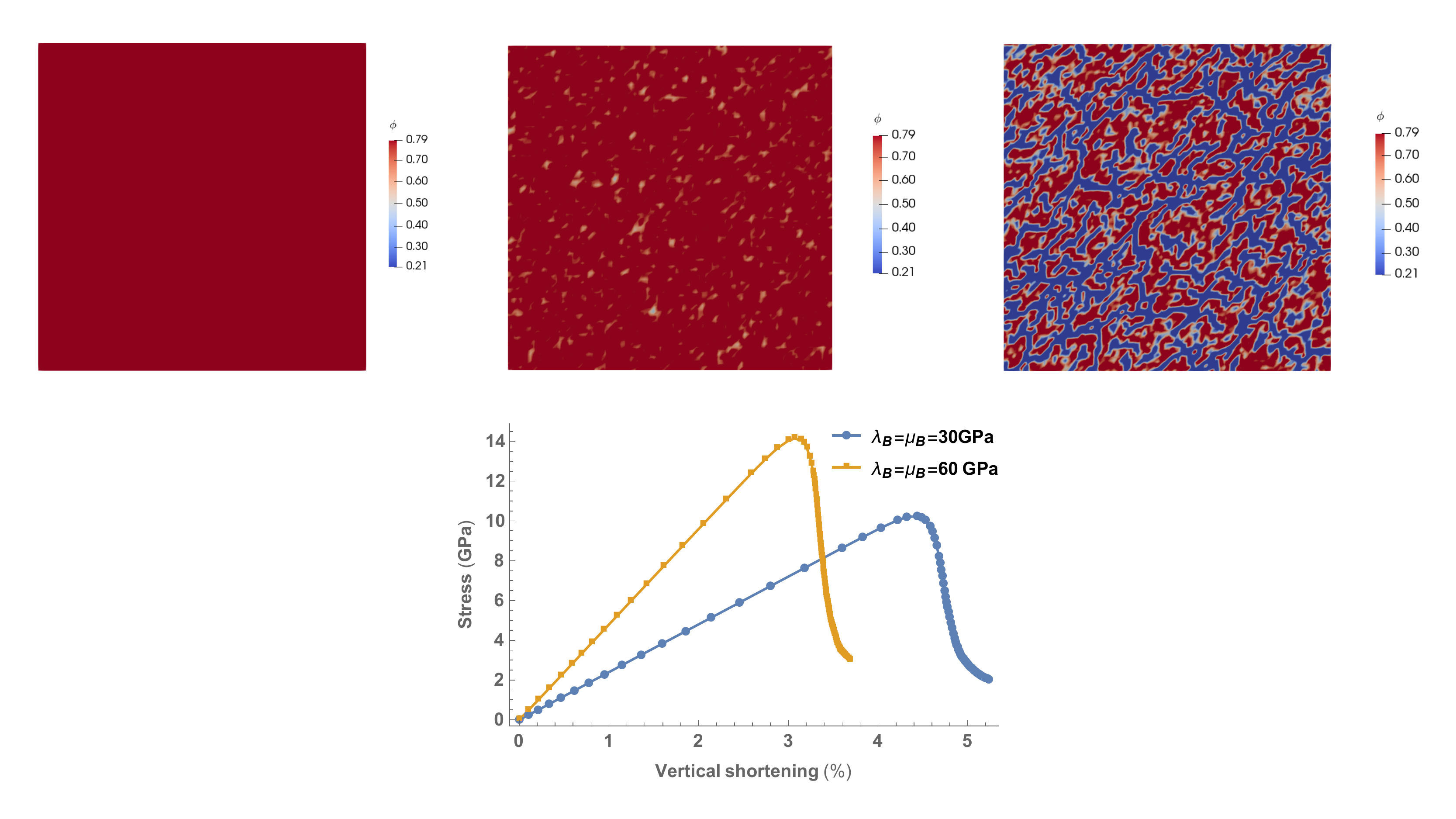}
\caption{Cracks nucleation in solid phase under isotropic compression at initial time, onset of softening ($vs=4.48\%$) and final time ($vs=5.23\%$) ($\lambda_A=1 GPa$, $\mu_A=0$, $\lambda_B=\mu_B=30 GPa$ and $\lambda_B=\mu_B=60 GPa$) with a mesh of $100*100$ triangular elements
  and associated stress/shortening curves measured on top boundary (in orange a material twice as stiffer as the one in blue.}
\end{figure}


We observe the nucleation of the blue weak phase under mechanical loading, result of the conversion of the initial fluctuations transferred form phase B to phase A I(see fig.2). It is interesting to look at the mechanical response associated to the phase change. It corresponds to the onset of mechanical softening (decrease of stress for increase of displacement). Indeed the apparition of the weak phase A accelerates the compression of the material. As expected, since the elastic moduli determine the tilt of the double-well under mechanical loading, the stiffer the material (orange curve) the faster the phase change and the onset of softening.

\subsection{Role of $\mu$: phase change CI}


We perform an oedometric compression of the REV of two microstructures, from the most basic to a more realistic one. Following our postulate that the Laplacian rate term's CI effects operate via controlling the variations of the interfaces curvatures, we apply our model first on a circle, i.e. a constant-curvature shape. In the following part we will upscale our study to geomaterials' CT scans.

\subsubsection{Circle-shaped inclusion}

We perform similarly a displacement-controlled compression of a circle-shaped weak phase A, captured in the strong phase B, this time in oedometric conditions (only the top boundary can move). This can represent the compression of a pore. Thus the initial conditions are as follows:\\

\begin{figure}[h!]
\centering
\includegraphics[scale=0.2]{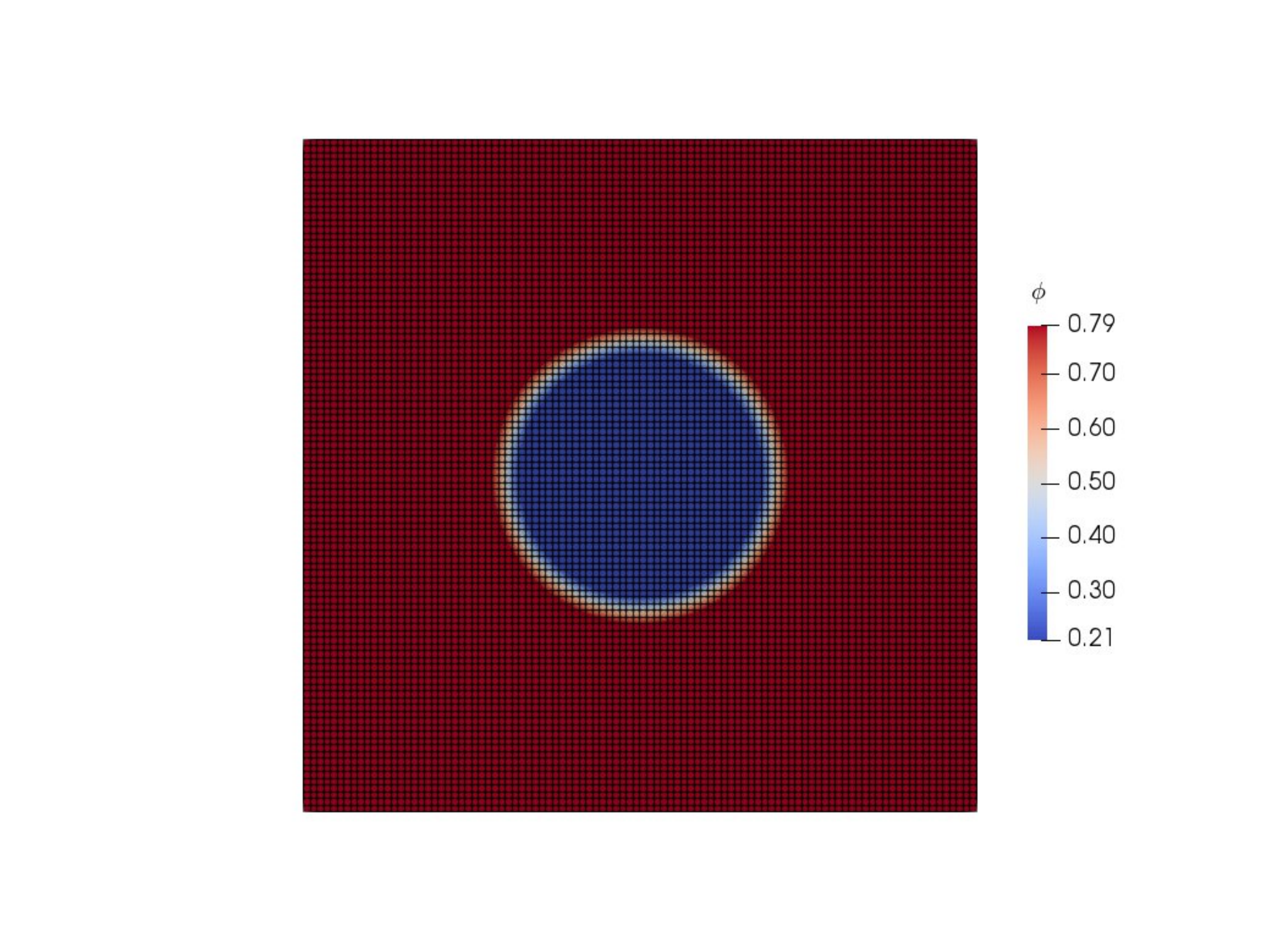}
\caption{Initial conditions for the oedometric compression of the circle-shaped inclusion (mesh with $100*100$ quadrilateral elements)}
\end{figure}

As per the previous dimensionless form of the equation, the model is fully parametrized by choosing $\mu$ and $\chi$. $\chi$ is fixed by choosing the elastic moduli $\lambda_A=1$, $\mu_A=0$ for the weak phase and $\lambda_B=30$, $\mu_B=30$ for the strong phase, and $G=1$. $/mu$ will be varied to study its influence. We visualize its influence through the output aspect, especially the interface curvatures, and the stress/strain curves. The following results are obtained by varying $\mu$ from $0$,$1$,$10$.


\begin{figure}[h!]
\centering
\includegraphics[scale=0.4]{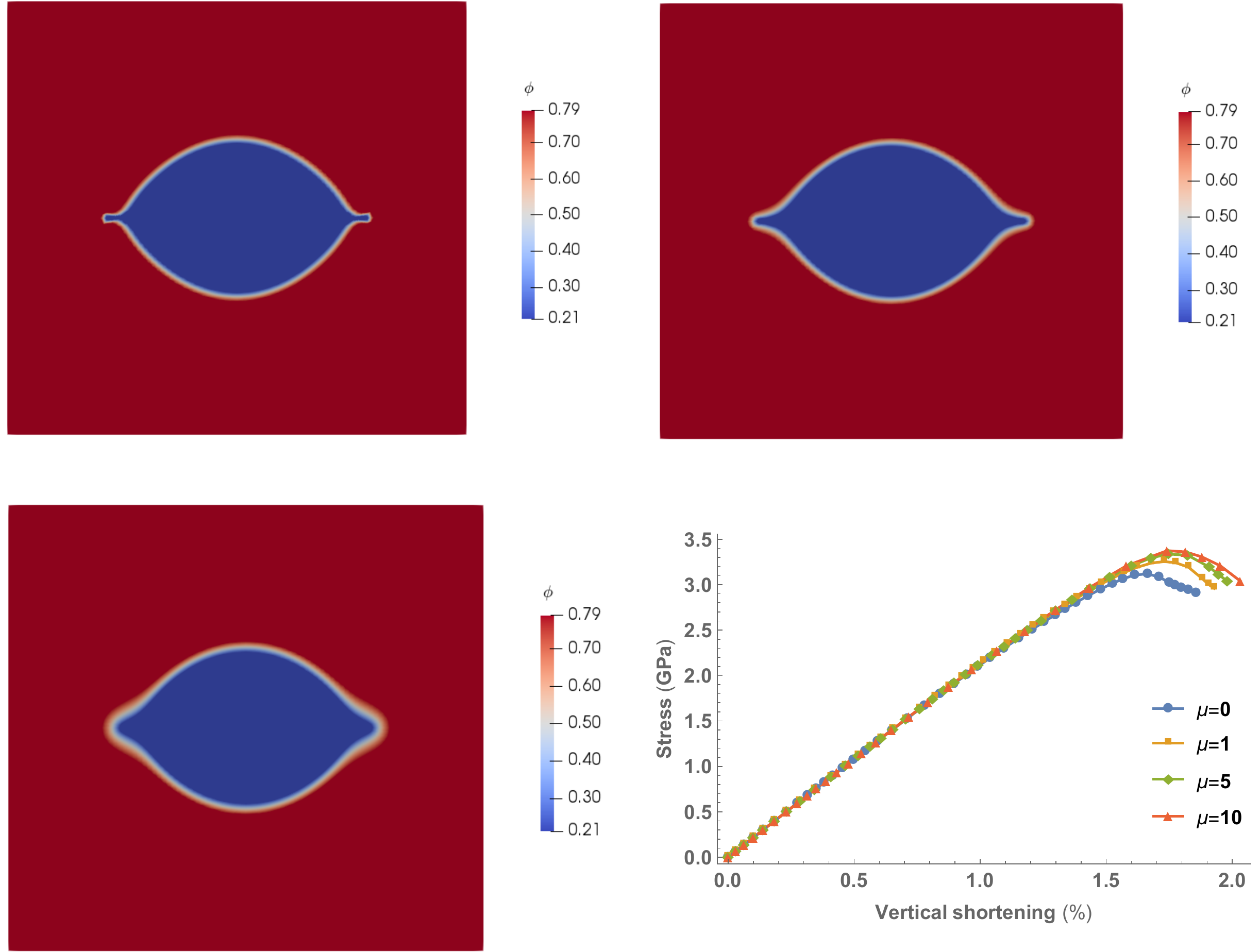}
\caption{Oedometric compression of a circle inclusion of weak phase (blue), just after softening for different values of $\mu$ (from left to right $\mu=0$,$\mu=1$,$\mu=5$), and associated stress/strain curves showing top-right translation with increasing $\mu$.}
\end{figure}

\subsection{Role of the free energy endothermic term: phase change inhibition}

In this last benchmark, we present the effect of the chemical coupling term $\beta^* c h'(\phi)$ that allows the production of the strong phase B ($\phi=1$). As explained previously, this chemical coupling term allows the change of sign of $f(\phi,\mat\epsilon,c)$ (and hence of $\dot\phi$). This can be visualize again the tilting of the double well, this time in the other way compared with the weak phase production (part 3.2.).

\begin{figure}[h!]
\centering
\includegraphics[scale=0.7]{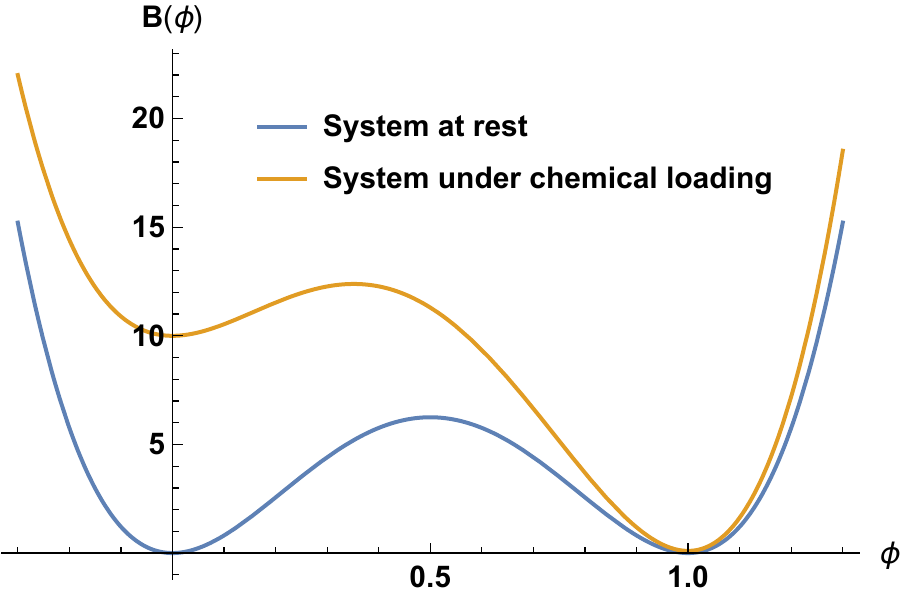}
\caption{Graph of the double-well potential $Gg(\phi) \approx B(\phi)$ with $G=10$ in blue and of the tilted double-well potential $B(\phi,\epsilon)=Gg(\phi)+\bar{H}(\epsilon,\phi)$ with $G=10$, $H_A(\epsilon)=0.1$ and $H_B(\epsilon)=10$ in orange}
\end{figure}

\subsubsection{Setup}

The initial setup consists of two half-circles separated by a thin layer of weak phase (representing the pores fluid) under a constant stress of $200MPa$ and oedometric conditions:

\begin{figure}[h!]
\centering
\includegraphics[scale=0.2]{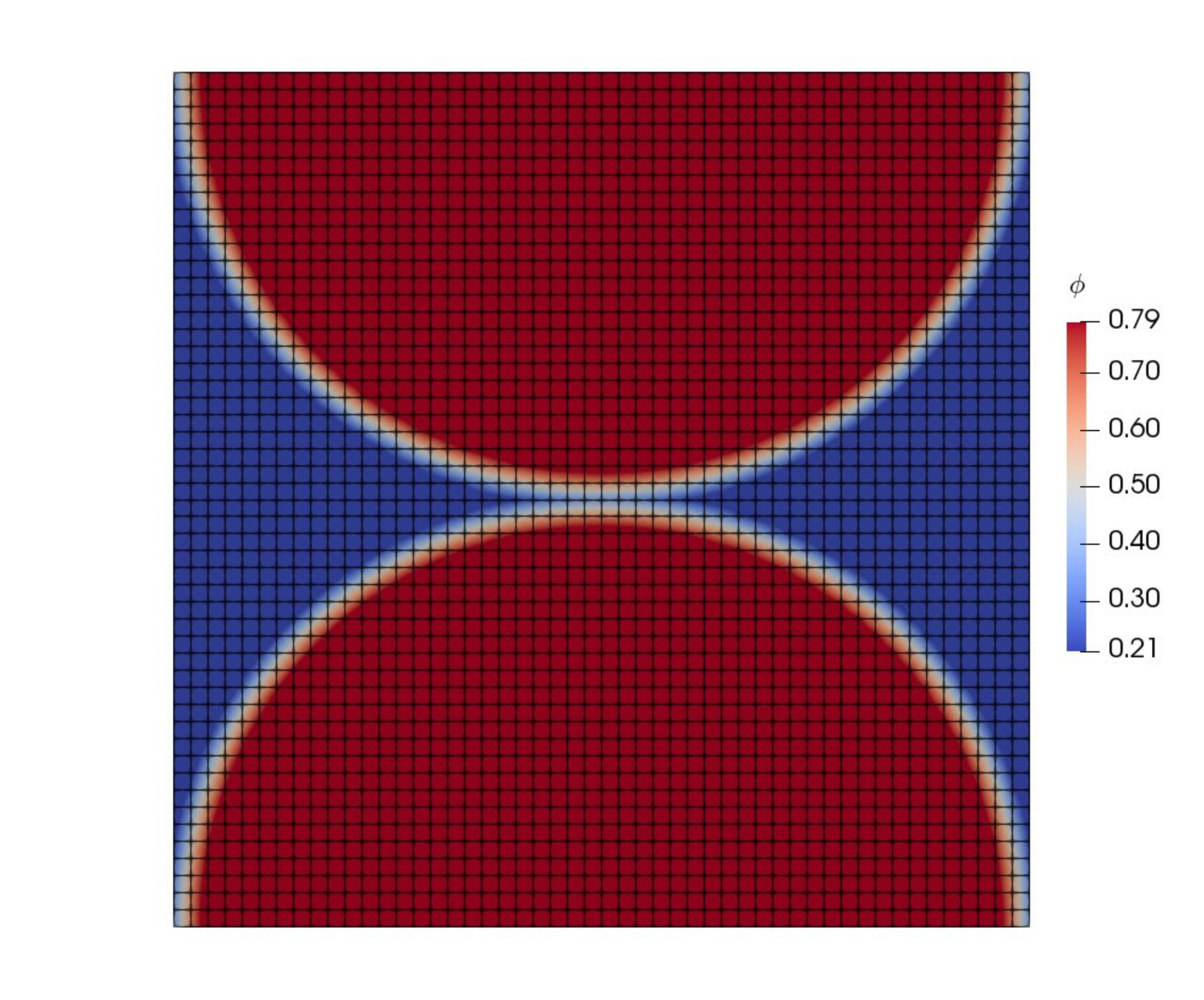}
\caption{Initial conditions for the benchmark of PSC model: two ideal spherical grains separated by a thin film of "fluid"}
\end{figure}

Whereas the thickness of the fluid film in between the grains should be reportedly of maximum few $nm$ \cite{Renard1997}, it is limited in our simulation by the mesh resolution. Indeed a quick calculation shows that our film thickness is approximately $12 \mu m$ ($grain \ diameter \approx 300 \mu m$, for a $50*50$ mesh elements, 2-element thick film). In order to have 1 mesh element measure say $1nm$, one should have a mesh of $3.10^5*3.10^5$ elements ($300/0.001=3.10^5$), which is clearly numerically unrealistic. We use the set of parameters $\alpha=0.1$, $\lambda_A=1$,$\mu_A=0$,$\lambda_B=\mu_B=30$, $\tau_4=\tau_1=1$, $D^*=10$. As explained in more details in the the next part, we consider two solid grains (geomaterials e.g.) separated by a shear-free much more deformable solid representing the liquid pores phase. 

Keeping that in mind, we perform different simulations with varying $\beta$ (see fig.8). It seems that the higher its value, the higher the chemical coupling (i.e. precipitation rate), translating into a slower compression (rightward translation of the displacement vs time curve. It is not clear how to quantify $\beta$ but we use the following rule of thumb: $\beta$ should be high enough to counterbalance the mechanically-induced dissolution (i.e. allow the reverse tilting of the double well); $\beta$ should not be too high to preserve coherent values of concentration $c$ as it appeared in the simulations. In that sense, we choose $\beta=0.05$ in the present case for instance. Indeed we observe in fig.8 that for too low values of $\beta$ (say $\beta<0.01$) the system's response is close to the case without chemical coupling ($\beta=0$).

\begin{figure}[h!]
\centering
\includegraphics[scale=0.5]{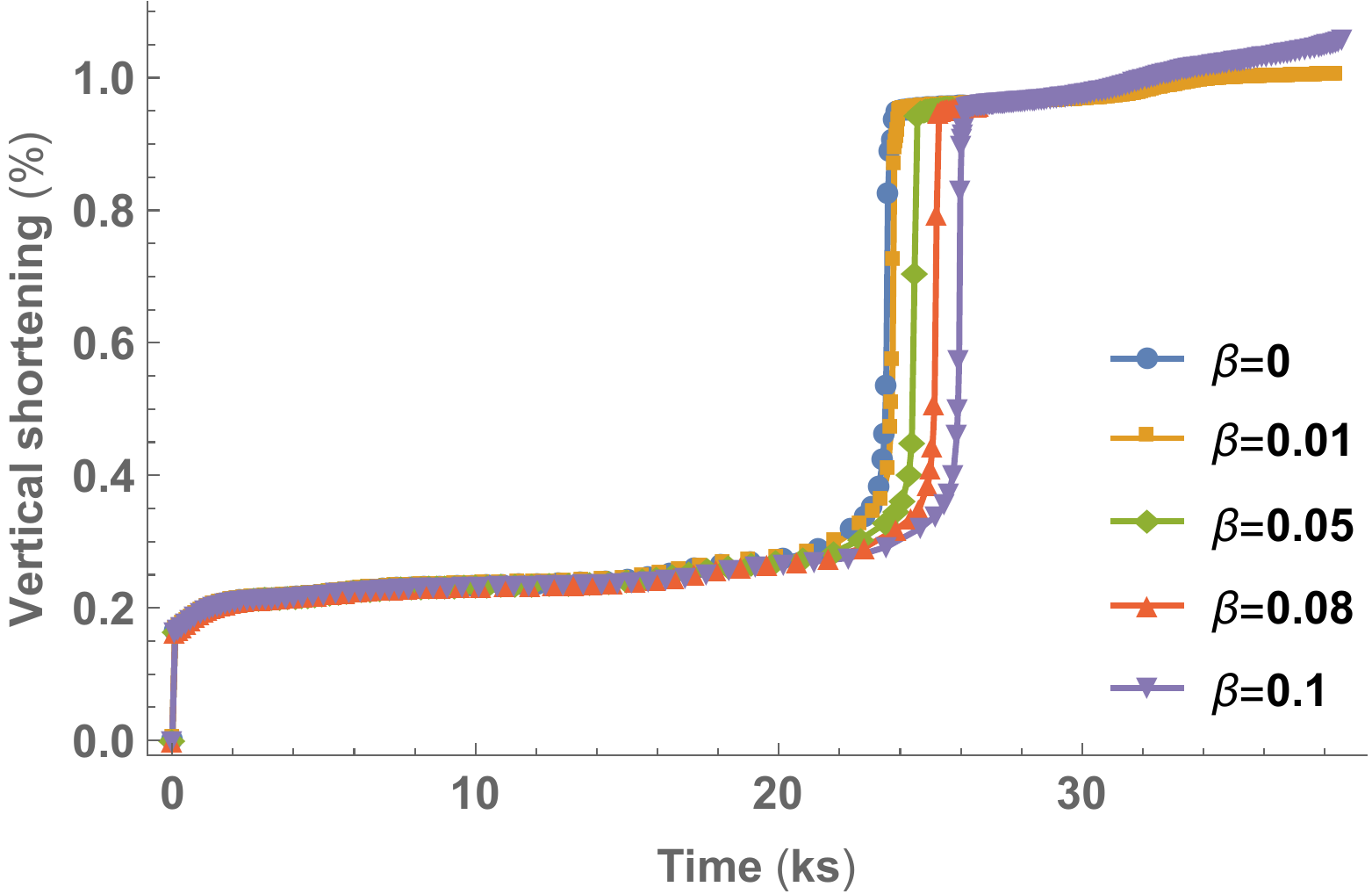}
\caption{Influence of $\beta$ on the the system's mechanical response: the higher its value the later the failure}
\end{figure}

\subsubsection{Mesh dependency}

Although the gross mesh $50*50$ allows to clearly visualize the dissolution at the contact zone, it seems that the jump in displacement is mostly dependent on the mesh resolution and therefore does not necessarily bear a physical meaning. We shall thus check the mesh dependency. We measure the time of the jump for finer and finer meshes. It appears that the finer the mesh the less accentuated the jump. In the light of the mesh convergence, we can choose a mesh of $100*100$ elements. 

\begin{figure}[h!]
\centering
\includegraphics[scale=0.5]{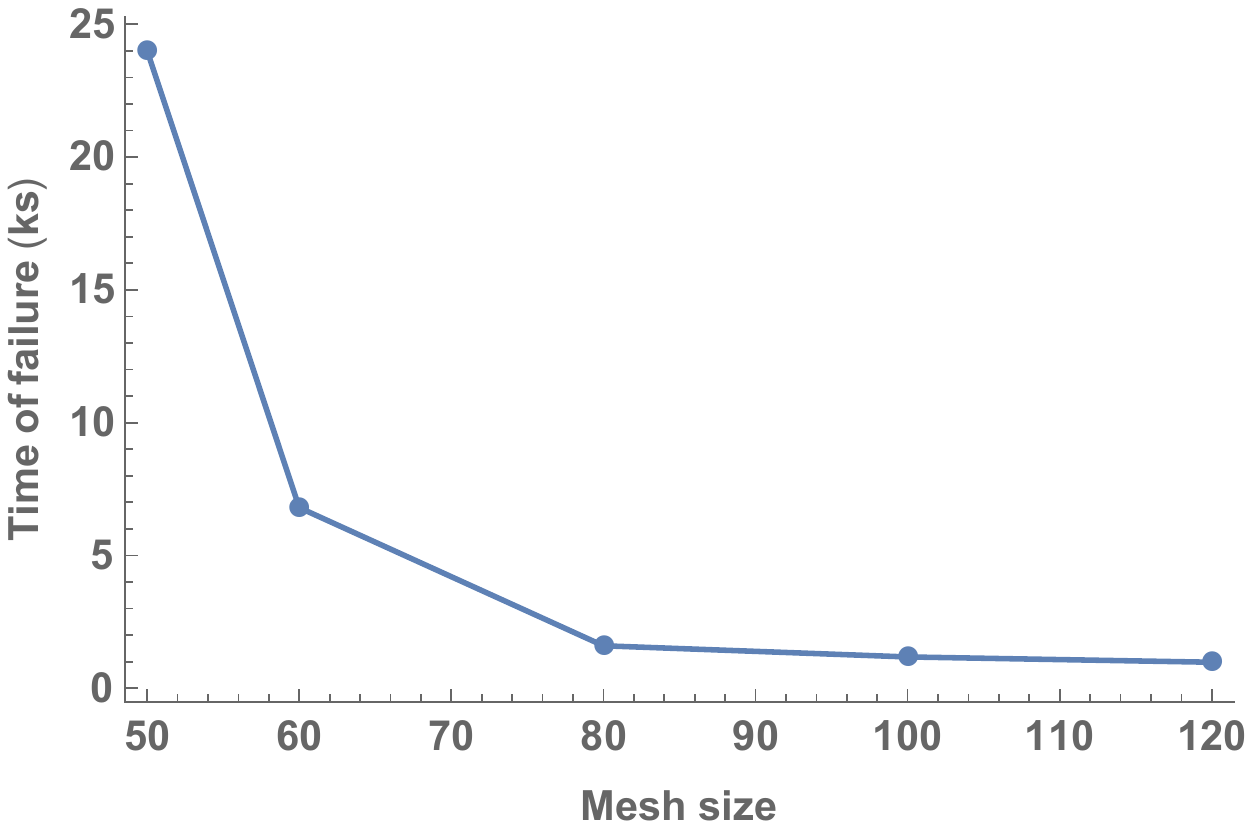}
\caption{Mesh convergence: measured failure time converges approximately for a mesh with at least $100*100$ elements}
\end{figure}

However, it appears that even for a converged mesh, the jump persists. We can attribute it to the discreet nature of our numerical modeling (FEM). Indeed, there will always be a last pair of elements (one from each grain) facing each other to be dissolved before the two flattened surfaces hit each other. In reality, we expect the dissolution to be smoother or at least that the jump induced by the last micro-particle would not be perceptible at the experimental scale. A crucial difference is also that in reality the microstructure is an assembly of grains that can ensure load restoration for each others when a contact between two grains gets flattened. 
Nonetheless, a possible physical meaning can still be attributed to this jump, viz. that this quasi-static phase before the jump is a sign of tertiary creep where the system does not receive enough energy at first to dissolve continuously the contact zone. 

\subsubsection{Numerical description of the PSC process}

Let us show the details of the process as they appear in our simulations for a given set of parameters, in particular $\beta=0.05$. We keep a rough mesh of $50*50$ to have a better visualization of the processes at stake, knowing that the mesh should be at least $100*100$ - then the jump in vertical shortening (failure) appears earlier but the output is qualitatively the same.
Our pressure solution model for the ideal case of two grains displays two clear stages: (1) the dissolution of the high-strain zone and (2) contact between the two flattened grains (see numerical results below in fig.9). The rounded grains surfaces are dissolved at the contact zone from the sides to the center until they become flat and there is no more support for the upper grain. Then a failure phase is characterized by a sudden increase in vertical displacement and the two flattened grains come into contact, still separated by a thin layer of fluid. Successive dissolution stages can be triggered on the flattened surface should the system be given enough energy (i.e. time in the present case of constant loading). One could associate those two stages respectively to the island-channel (IC) representation (\cite{Dysthe2003}, first in \cite{Raj1981}) and the thin-film (TF) model (\cite{Weyl1959}).

Now let us show the details of the process as they appear in our simulations for a given set of parameters.

\begin{figure}[h!]
\centering
\includegraphics[scale=0.4]{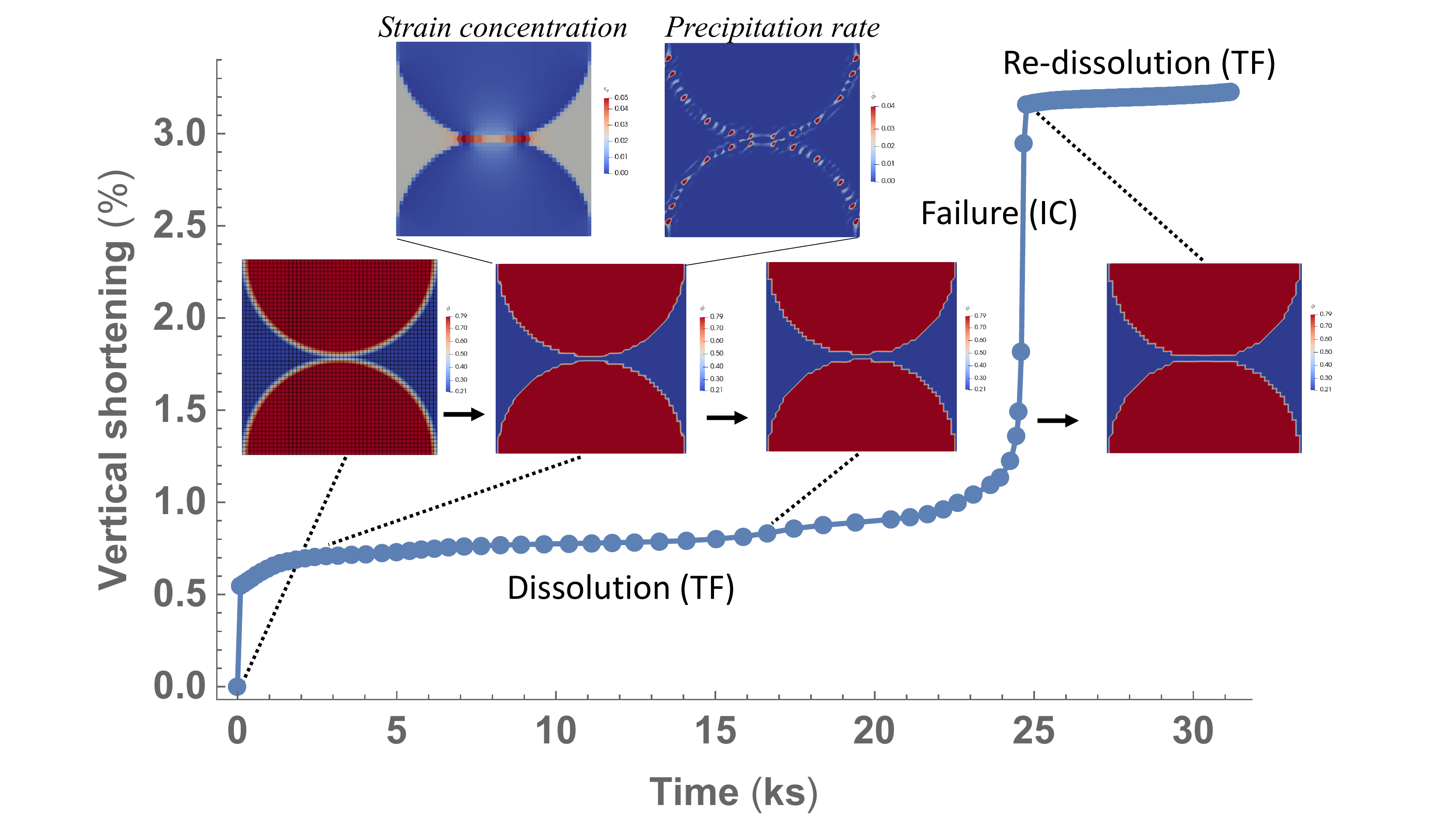}
\caption{PSC of two-grain benchmark at during dissolution until the rounded surfaces get flattened, and failure occurs at $t=24.3 ks$. Then the new contact surfaces are available for a new stage of dissolution. The concentration of volumetric strain and the precipitation rate (i.e. $\dot\phi_+$ are shown at $t=3.43 ks$. Those numerical images should be related to the green curve in fig.8.}
\end{figure}

We will see in part 4.2. that the present benchmarks conclusions are not as ideal for an irregular microstructure, since the response depends on the specific MG.

\section{Results for chemo-mechanical degradation of microstructures}

After benchmarking the different features of our model, we apply it to more realistic microstructures, those of geomaterials. We will use as input digitalized CT scans images of a sandpack obtained from a sandpack's CT scan from \cite{Dong2009}. We refer to the website https://www.imperial.ac.uk/earth-science/research/research-groups/perm/research/pore-scale-modelling/micro-ct-images-and-networks/sand-pack-lv60a. Unless mentioned otherwise, we use the layer 159 therein and consider a 2D problem.

\subsection{Digital modeling of geomaterials}

We consider now the two phases to be the rock matrix or grains (phase B, $\phi=1$) and the pores (phase A, $\phi=0$). As a first approximation, we model the pores phase a solid as well, similarly to \cite{Kassner2001}. As such, the pores fluid (air and/or liquid) will be then taken as a shear-free solid much more deformable than the matrix phase. Ideally the mechanics should be coupled with hydrodynamics but we will restrict our model to this assumption for now. This implies the elastic moduli to fulfil $\mu^A=0$ and $\lambda^A \ll \lambda^B$. Note that then $\lambda_A=K_A$ with $K$ the bulk modulus. The elastic energy of each phase reads now, in 2D (plane strain):

\begin{equation}
H_A=\frac{1}{2}\lambda_A({\epsilon_{xx}^{A}}^2+{\epsilon_{yy}^{A}}^2)
\end{equation}

\begin{equation}
H_B=\frac{1}{2}\lambda_B({\epsilon_{xx}^{B}}^2+{\epsilon_{yy}^{B}}^2)+\mu_B({\epsilon_{xx}^{B}}^2+{\epsilon_{yy}^{B}}^2+2{\epsilon_{xy}^{B}}^2)
\end{equation}

We remind that we choose to work in the following consistent set of units: mm, ks, J, GPa, kg.\\

Let us consider the pores phase A as saturated with water and then $\lambda_A=K_A \approx 1 GPa$ (and $\mu_A=0$). As for the grains phase B, we cannot consider the mechanical characteristics of sand or sandstone as such but independently from the pores. In that sense, the grains phase can be considered as a rock with very low porosity, like granite. Hence we choose $\lambda_B \approx 30 GPa$ and $\mu_B \approx 30 GPa$, which are values close to what can be found in standard literature. We apply an oedimetric displacement-controlled compression directly proportional to the time $1*t$.

The initial conditions are obtained by digitalizing the ct-scan binary image (left) into an image usable numerically in MOOSE (using the function ImageReader). We use a mesh $70*70$ for a initial ctscan whose resolution is $300*300$. Obviously, should the resolution be preserved, we should use a mesh of $300*300$, but this is unnecessarily computationally expensive for our qualitative study. However, to model the exact microstructure and obtain realistic quantitative results one should use a $300*300$ mesh. The dimensions of the ct-scan are $3*3*3 mm^3$. Therefore, the reference length for the ct-scan simulations will be $l_0=3 mm$. As for the reference length for the benchmarks of at the grains scale, we will take $l_0=0.3 mm$ e.g., obtained from the granulometry of the same sand available in \cite{Talabi2009}.

\begin{figure}[h!]
\centering
  \includegraphics[width=0.7\linewidth]{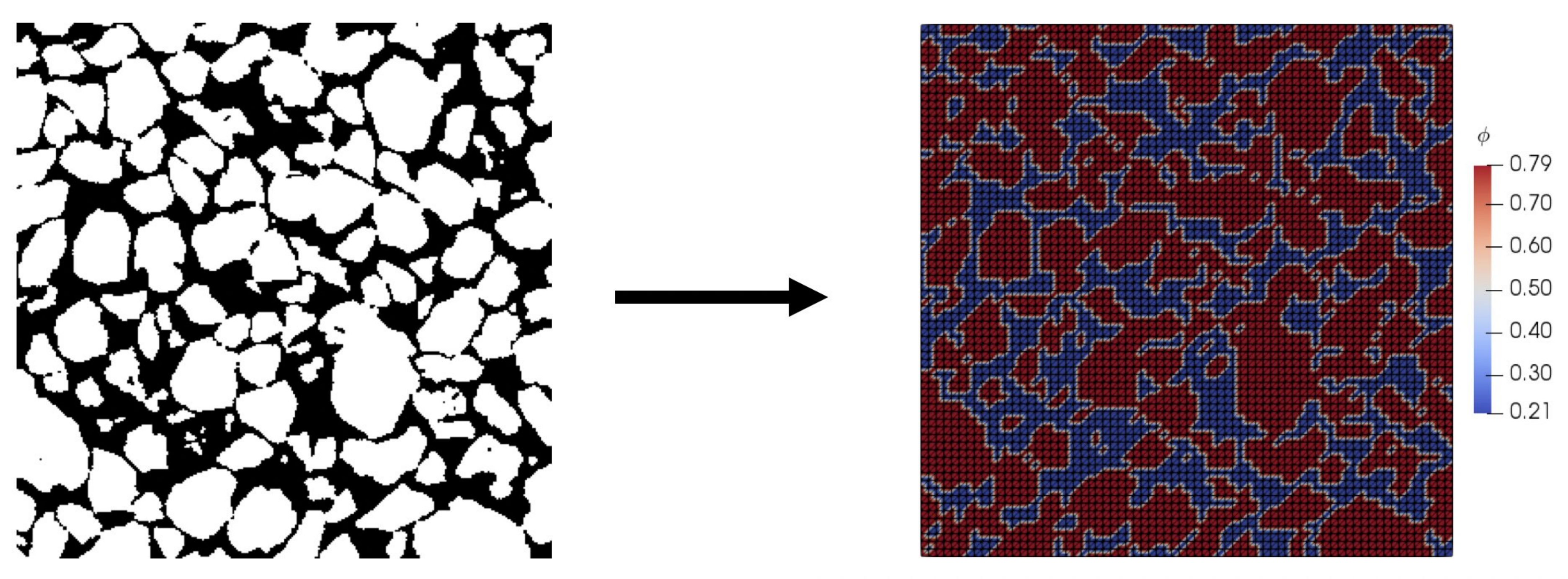}
   \caption{Digitalization of a LV60A sandpack's CT scan obtained from \cite{Dong2009}}
\end{figure}

The parameter $\alpha$ contains the squared length scale $l_0^2$ and therefore should be divided by $100$ as compared with the simulations on grains since for grains $l_0=0.3mm$ and for a CT scan $l_0=3mm$ (assuming the same material). So we should have here $\alpha=0.001$ but as mentioned before when $\alpha$ is too small, the numerical results are not satisfying. In particular, the values of $\phi$ get too much out of the range $[0,1]$. We thus stick to $\alpha=0.1$ (and $\alpha=0.01$ in part 4.2.

We now look at the influence of our new coefficient $\mu$ (more exactly $\tau_2$), the phase-field viscosity. As we assumed previously and shown analytically in \cite{Guevel2019a}, the coefficient multiplying $\Delta\dot\phi$ characterizes the change of interfaces curvature and the phase change kinetics, and more generally the convergence to equilibrium. It makes then sense to associate this term with phase changes inducing change of interface curvature, which are a priori all non purely volumetric changes, like most degradation processes (dissolution e.g.). We verify this postulation by comparing the configuration of the present CT scan at the same time of deformation for different values of $\mu$:


\begin{figure}[h!]
\centering
  \includegraphics[width=1.2\linewidth]{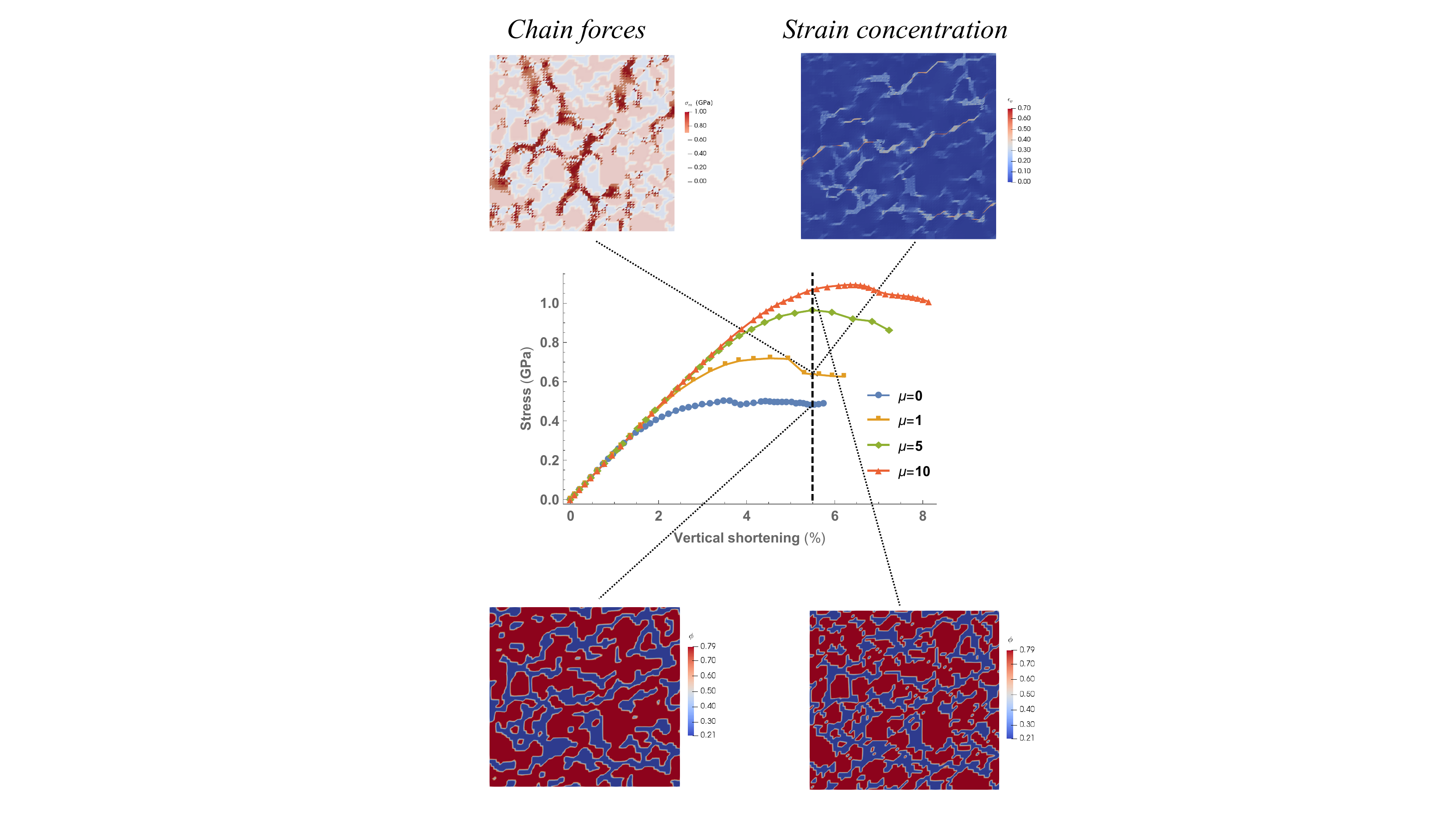}
   \caption{Simulations outputs for CT scans simulations for a same vertical shortening of $16.49\%$ for different values of $\mu$ ($\mu=0$ bottom left, $\mu=10$ bottom right)
  and (b) associated stress/vs curves, dotted vertical line placed at $vs=16.49\%$}
\end{figure}


Visually, as expected, increasing values of $\mu$ delay the change of curvatures, and as a result for low $\mu$ the grains appear more "mixed" than higher values. In terms of mechanical response, $\mu$ controls the onset of softening, i.e. decrease in stress for increase in strain. Thus onset of softening indeed corresponds to microstructural phase change, as it is usually assumed.\\

We conclude this discussion on the role of $\mu$ to show its implications on the dissipation, quantity at the heart of our model's derivation and reasoning \cite{Guevel2019a}. We display here below the dissipation $D=D_n+D_t=\tau_1 \dot\phi^2 + \tau_2||\nabla\dot\phi||^2$ (integrated over the digitalized CT scan domain) for $\mu=0$ (left) and $\mu=1$ (right). We associate the maximum mean stress calculated on the same domain. A drop in the maximum mean stress (softening) corresponds to a major phase change. Those dropped are accompanied by a peak of dissipation.


\begin{figure}[h!]
\centering
  \includegraphics[width=1\linewidth]{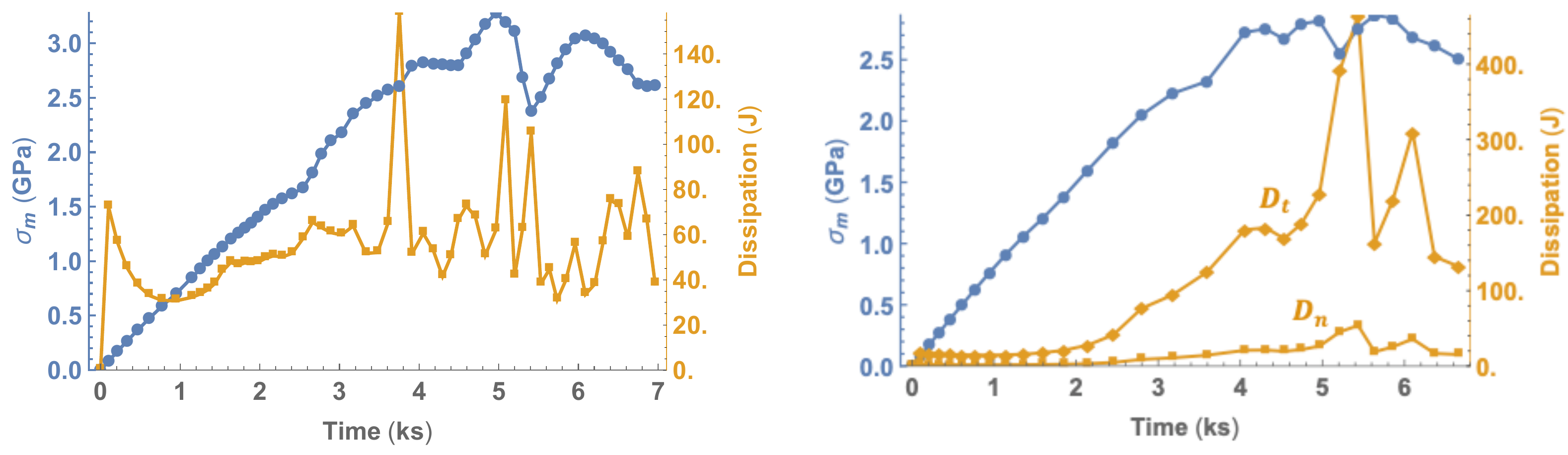}
   \caption{Maximum mean stress and dissipation for $\mu=0$ and $\mu=1$. The dissipation has two components $D=D_n+D_t=\tau_1 \dot\phi^2 + \tau_2||\nabla\dot\phi||^2$ ($D_t=0$ for the case $\mu=0$).}
\end{figure}

Note that the dissipation components are calculated assuming $\tau_1=1$ (as everywhere in the present numerical simulations) and $\tau_2=\mu \tau_1 l_0^2$.
Interestingly, the evolution of the system's dissipation is significantly different depending if the Laplacian rate term is activated or not. In the latter case ($\mu=0$), the dissipation is more irregular whereas in the former case ($\mu=1$) the dissipation evolution consists in two peaks, mostly due to the tangential component, corresponding to the significant phase changes of the process. \\

Let us know combine the previous model with chemical coupling and thus get some insights on PSC.

%
%
%
%
%
%
%
%
%
%

 \subsection{Application to PSC}

%

Since we are using in this work a sand pack as CT scan input, let us fix the pressure solution reaction by considering the dissolution reaction of silica (${SiO_2}_{(s)} + {2H_2 O}_{(l)} \rightleftarrows {Si(OH)_4}_{(aq)}$). Thus the grain phase is silica, the pore phase is water and the solute (of concentration $c$) is silicic acid. However, as it is the case in the present work, we focus on a preliminary qualitative understanding of the processes rather than obtaining quantitative estimates.

\subsubsection{Parameters setup}

Let us first choose the parameter $\beta$. As observed in part 3.4, it seems that the higher its value, the higher the precipitation rate, translating into a slower compression (rightward translation of the displacement vs time curve. It is not clear how to quantify $\beta$ but we use the following rule of thumb: $\beta$ should be high enough to counterbalance the mechanically-induced dissolution (i.e. allow the reverse tilting of the double well); $\beta$ should not be too high to preserve coherent values of concentration $c$ as it appeared in the simulations. In that sense, we choose $\beta=0.1$ in the present case. We can see that $\beta$ has the anticipated effect, but not as clearly as in the benchmark case for two ideal grains. Indeed we expect the MG to have a significant effect in the ctscans simulations. 

\begin{figure}[h!]
\centering
\includegraphics[scale=0.5]{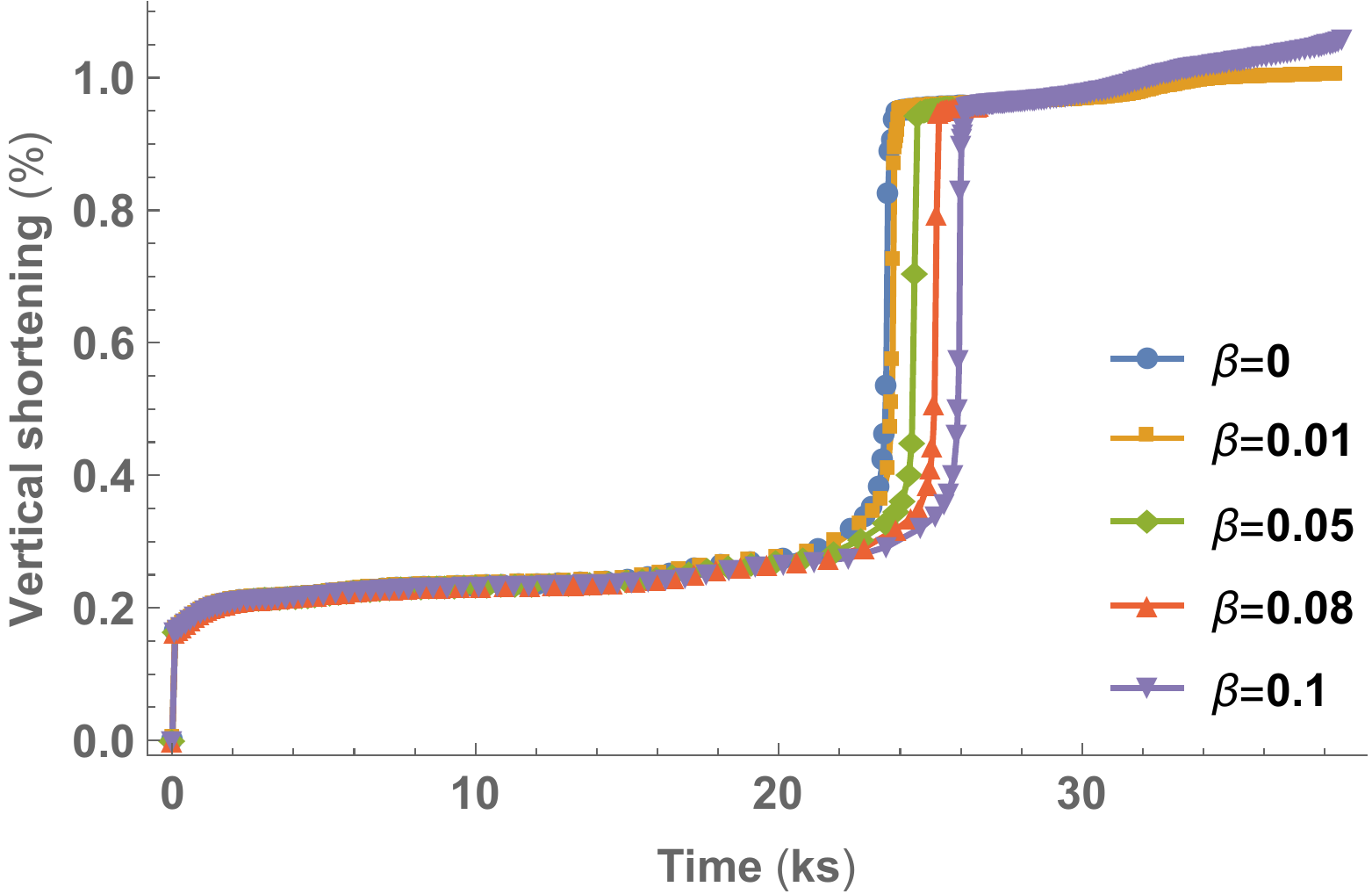}
\caption{Influence of $\beta$ on the microstructure's response: the higher its value the slower the compression. However this is less obvious for the present microstructure than for the ideal two-grain benchmark}
\end{figure}

Then we choose $\alpha=0.01$ and $D^*=0.1$ to keep $\alpha<D^*$.

\subsubsection{Chemo-mechanical response}

As expected from the model's equations, dissolution should happen in high stress/strain zones ($\hat\chi(\mat\epsilon,c)>0$) and conversely precipitation should happen in low stress/strain zones ($\hat\chi(\mat\epsilon,c)<0$). The numerical results below are shown at $t=3.41$ at the end of a major phase change, i.e. jump in vertical shortening (cf fig.14), the dissolution of a supporting bridge in the center of the ctscan (to be compared with initial state in fig.11). To illustrate the pressure dissolution process, the volumetric strain, solute concentration and order parameter rate are displayed as well. As observed previously in the case without chemical coupling, dissolution is favored in the strain localization zones (negative values of $\dot\phi$ in bottom right corner picture), accompanied by the production of solute (red zones in bottom left corner picture). In addition, in the present case of chemo-mechanical coupling, the solute is allowed to precipitate in the low-strain zones (dark red dots in bottom right corner picture). However, it is not always clear whether the production of solid phase ($\dot]phi>0$) is due to precipitation (due to the term $\hat\chi(\mat\epsilon,c)$ when negative) or grain boundary diffusion (due to the term $\alpha\Delta\phi$). This is why one should be careful not to choose $\alpha$ too large. We assume that the "red hots", corresponding to the pores closure (or pores collapse) are mostly due to grain boundary diffusion. Nonetheless, the central zone we are focusing on (black circle) clearly displays precipitation and not closure. The solute available after dissolution (red zone in center of bottom left corner picture) precipitates on the wall of the pore (zoom-in picture). This seems favored by the highly localized strain (see black circle in top right corner picture), allowing nearby precipitation as the strain quickly decreases around. The production of solid phase most likely results froma combination of grain boundary diffusion and precipitation.

\begin{figure}[h!]
\centering
\includegraphics[scale=0.4]{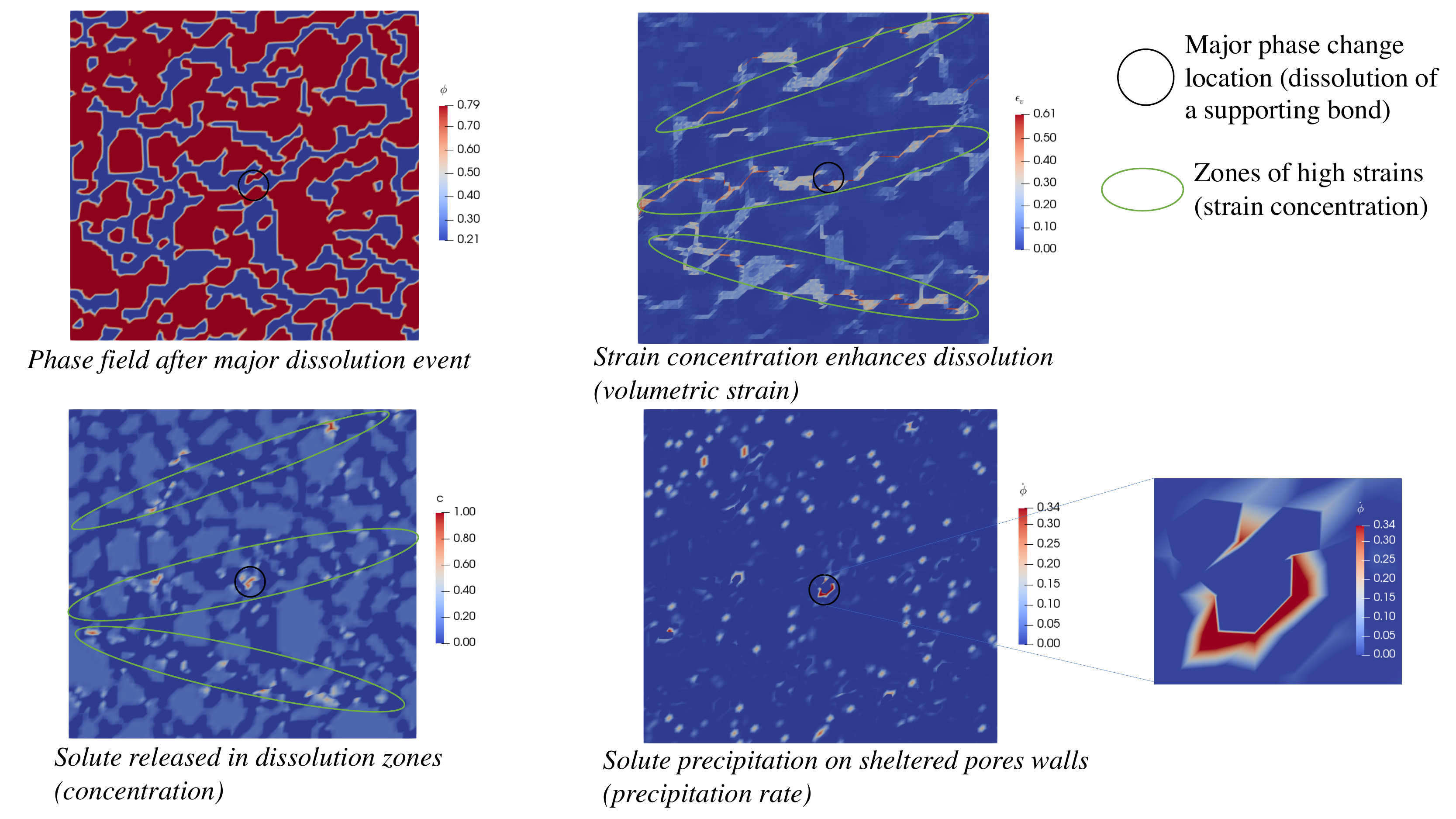}
\caption{Visualization of the microstructure's state at $t=3.41 ks$ at the end of the major phase change. Top right: order parameter. Top left: volumetric strain. Bottom right: solute concentration. Bottom left: order parameter rate. Zoom in: only positive values of order parameter rate, showing the dissolution/precipitation interaction. The particular MG drives the strain concentration, which drives the dissolution/precipitation.
}
\end{figure}

\subsubsection{Influence of the microstructure's geometry (MG)}

Since PSC results from a stress-induced mass transport, the primordial process is stress/strain concentration, which depends strongly on the MG. We therefore expect to have a different response for different geometries. The results below are obtained from different layers of the same digitalized sand pack used previously \cite{Dong2009}. As expected, even though from the same specimen of geomaterial, the response is significantly different from different parts of the specimen. Since we can visualize the evolution of the MG thanks to PFM, we can observe the grains evolution and its correspondence to the stress/strain curves. For instance, the geometry of MG1 is less prone to stress/strain concentration and thus exhibits a longer phase of strain hardening in order to load more energy, correspond to grains reorganization. MG2 displays a faster phase change (at $t \approx 4 ks$ vs $t \approx 7.5 ks$), having an initial MG more favorable to strain concentration than grain reorganization.

\begin{figure}[h!]
\centering
\includegraphics[scale=0.7]{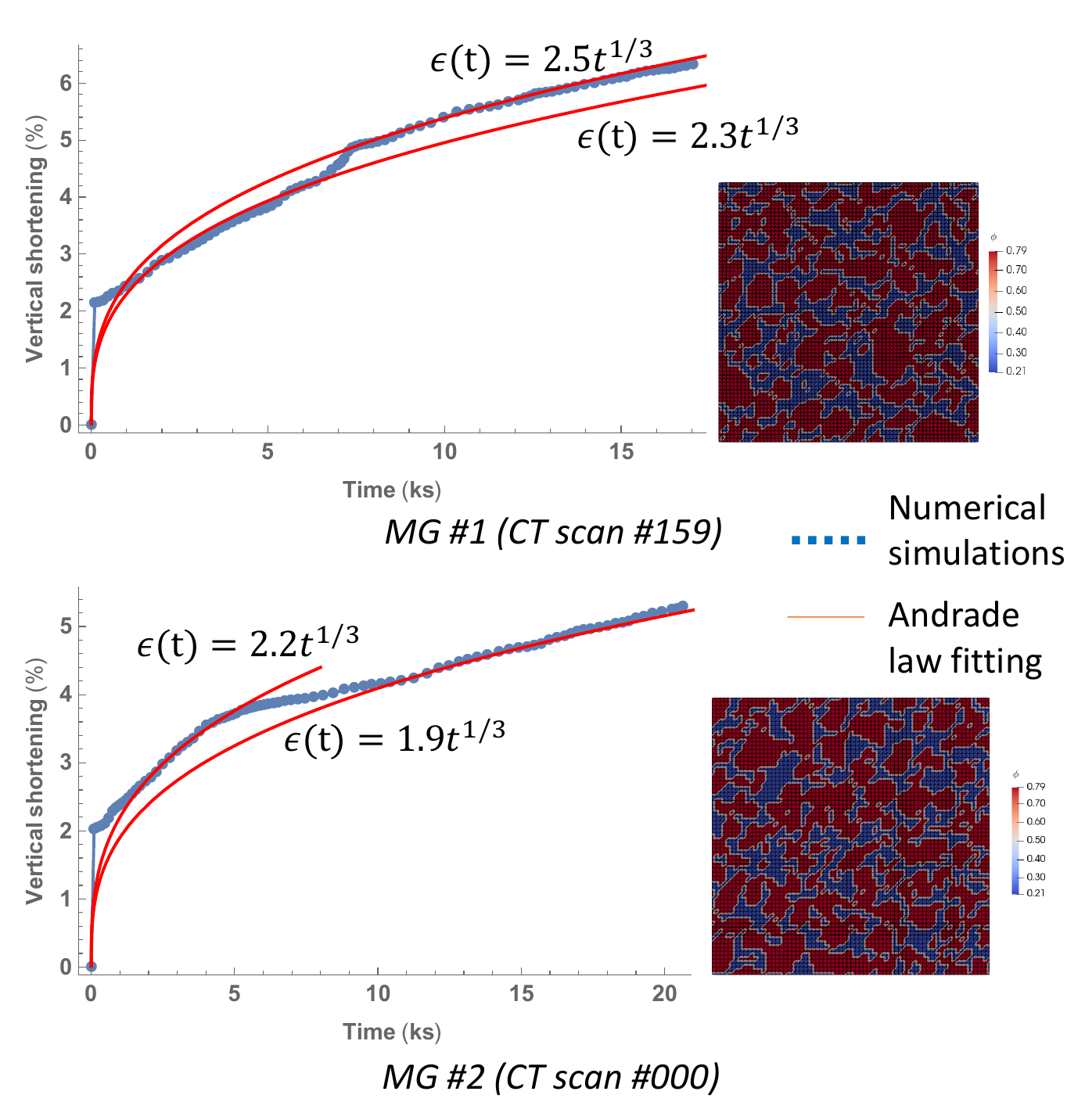}
\caption{Dynamic Andrade creep laws ($\epsilon(t) \sim t^{1/3}$), separated by weakening events (different MGs have different primitive processes)
}
\end{figure}

Furthermore, we fit power laws in the cubic root of time, the so-called Andrade creep law, with good agreement. The adequateness between Andrade creep (from metallurgy initially) and PSC has been shown in \cite{Dysthe2002} and \cite{Dysthe2003}, with good agreement with experiments on salt.

\subsubsection{A universal creep law?}

The investigation towards building a unique creep law has not found an end just yet. In particular, it is not clear which is the rate-limiting process. It seems however that significant progress has been achieved by analogy with metallurgy. It is argued in \cite{Dysthe2002} that the characteristic length scale of the contact between two grains in PSC may grow as the cubic root of time, similarly to the Andrade creep law. In a related work \cite{Dysthe2003}, it is infered, via ideal spherical geometries and constitutive description of the contact, that the vertical shortening may as well follow an Andrade creep law. A key argumentation in \cite{Dysthe2002,Dysthe2003} is the acknowledgement that PSC is a transient process, whence the use of dynamic contact laws. This is corroborated by high-precision experiments on salt. The vertical shortening seems to follow such law upon change in loading stress.

Interestingly, it seems that such Andrade fitting is relevant to our numerical results, even though we dropped the ideal spherical packing for a more accurate representation of the MG. However, we tend to see variations of the fitting from one Andrade law to another, even with a constant stress. The jump occurs upon weakening events, i.e. significant phase change. We would thus argue that more that a universal creep law for PSC, the response seems to follow an adaptative Andrade creep law, strongly dependent on the particular MG's dynamics. A variation of Andrade creep along the PSC process makes sense inasmuch as the fitting has been shown in the works above to be dependent among other things on the film layer thickness in between grains. This film layer obviously varies significantly in our (dynamic) results as grains reorganize and dissolve/reprecipitate.

\subsubsection{Physical meaning of $\mu$}

%
%
%
%
%

As showcased previously, the term in $\mu$ represents the phase-field viscosity, i.e. control the kinetics of the phase change (the higher its value the more delayed the process). It can encapsulate a priori any CI effects of the main process without having to model it explicitly. It can be associated with the activation energies $Q_i$ of the different catalytic effects $i$ in the form $\mu=\sum_{i} A_i e^{-\frac{Q_i}{k_B T}}$. For instance for PSC two main such effects can be the temperature \cite{Niemeijer2002} and the presence of clays \cite{Renard2001}, both assumed to enhance the process. We get the following responses for a fixed value of $\beta=0.1$:

\begin{figure}[h!]
\centering
\includegraphics[scale=0.4]{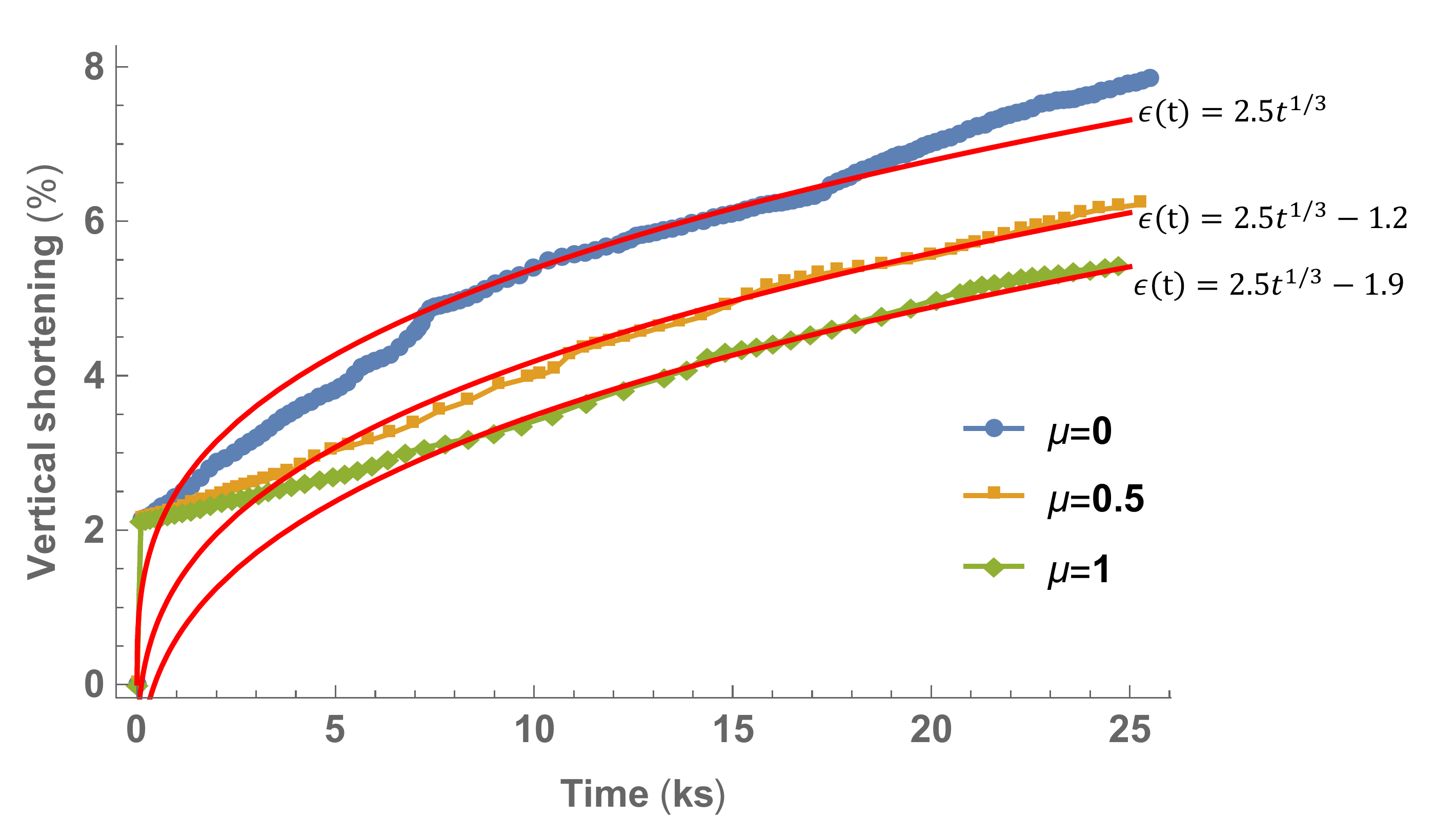}
\caption{Influence of $\mu$ on the microstructure's response: no qualitative change (grains rereorganization followed by major phase change) but delayed as $\mu$ increases}
\end{figure}

A variation in $\mu$ corresponds to a vertical translation of the vertical shortening (constant slope of Andrade creep fitting), similarly to an increase of temperature a priori \cite{Niemeijer2002} and clay content \cite{Renard2001}.

\section{Conclusion}

We have studied numerically the different features of our previously developed CPFM and applied to geomaterials' MGs undergoing chemo-mechanical degradation. Two main conclusions should be emphasized. Firstly, our new Laplacian rate term is shown to control the variations of the interfaces curvatures and as such acts as a CI for degradation processes. For PSC, that could correspond to temperature or clay content, both enhancing the process. Secondly, the tracking of the MG's dynamics thanks to PFM is proven to have a significant influence on the system's behavior at the upper scale. This is particularly interesting for PSC modeling, for which the MG modeling is usually restricted to ideal spherical packing. Our results corroborate the already existing observation that microstructurally-driven processes like PSC are the result of transient interacting instabilities at the grain scale, but also that a dynamic Andradre creep law seems to prevail for PSC. Our results provide thus preliminary insights towards understanding better the influence of the MG of a system's response.

%
%



\appendix


\bibliography{library}

\end{document}